# Correlation between nucleotide composition and folding energy of coding sequences with special attention to wobble bases.

Jan C. Biro[1]
Homulus Foundation – 612 S Flower Str., #1220; Los Angeles; 90017 CA., USA
Jan.biro@att.net
www.janbiro.com

[1]Corresponding author

## Abstract

**Background:** The secondary structure and complexity of mRNA influences its accessibility to regulatory molecules (proteins, micro-RNAs), its stability and its level of expression. The mobile elements of the RNA sequence, the wobble bases, are expected to regulate the formation of structures encompassing coding sequences.

**Results:** The sequence/folding energy (FE) relationship was studied by statistical, bioinformatic methods in 90 CDS containing 26,370 codons. I found that the FE (dG) associated with coding sequences is significant and negative (407 kcal/1000 bases, mean ± S.E.M.) indicating that these sequences are able to form structures. However, the FE has only a small free component, less than 10% of the total. The contribution of the 1st and 3rd codon bases to the FE is larger than the contribution of the 2nd (central) bases. It is possible to achieve a ~ 4-fold change in FE by altering the wobble bases in synonymous codons. The sequence/FE relationship can be described with a simple algorithm, and the total FE can be predicted solely from the sequence composition of the nucleic acid. The contributions of different synonymous codons to the FE are additive and one codon cannot replace another. The accumulated contributions of synonymous codons of an amino acid to the total folding energy of an mRNA is strongly correlated to the relative amount of that amino acid in the translated protein.

**Conclusion:** Synonymous codons are not interchangable with regard to their role in determining the mRNA FE and the relative amounts of amino acids in the translated protein, even if they are indistinguishable in respect of amino acid coding.

**Key words:** RNA structure, folding energy, wobble bases, codon

# Background

Messenger RNA was originally not expected to have any secondary structure, because it was simply supposed to pass through the ribosomes (as a magnetic tape passes a tape-recorder) [1] and any secondary structure was thought to interfere with the translation process. However, mRNA has considerable total folding energies (TFE) depending on the number and distribution of complementary bases (407 kcal/1000 bases, mean ± S.E.M., n=147). The TFE is the sum of two components. First, the compositional component (CFE), which is determined only by the numbers of the four bases and their relative proportions, is equal to the dG value of shuffled sequences. Second, there is the positional component (PFE), which is determined by the position of bases and can be less or more than the CFE. This component is called Folding Free Energy (FFE) and is the difference between the dG values of intact and shuffled RNAs. Most attention has been paid to the FFE because it is required to form a unique structure, while the CFE defines numerous equally possible structures. While the CFE is a measure of random complexity, the FFE is the measure of ordered, structural complexity.

The secondary structure and complexity of mRNA became an important issue because it influences the accessibility of the mRNA to regulatory molecules (proteins, microRNA), its stability and its level of expression. In addition, new theoretical considerations and experimental evidence suggest that mRNAs may even play role in carrying and providing structural information for translated proteins and might serve as chaperons.

It is suggested that mRNA secondary structures are conserved and play important roles in (a) splicing [2], (b) control of gene expression [3], (c) discontinuous translation and pauses in protein synthesis [4, 5], (d) determining the protein secondary structure [6-8] and (e) regulating gene expression level and accuracy [9].

The aim of this work is to study the relationship between mRNA primary and secondary structure, base composition and structural complexity (measured as folding energy). Special attention is paid to the role of wobble bases in modifying mRNA thermodynamics.

# Methods

Ninety coding sequences of proteins with known 3D structures were selected from the PDB. This selection contained 26,370 codons. Care was taken to avoid very similar structures in the selections. The propensity towards alpha helices was monitored during selection and structures with very high and very low alpha helix contents were also selected to ensure a wide range of structural representations. The possible influence of different kingdom (prokaryotes and eukaryotes) or variations in environmental conditions on mRNA folding stability were not considered in this study.

Single-stranded RNA molecules can form local secondary structures through the interactions of complementary segments. Watson-Crick (WC) base pair formation lowers the average free energy, d$G$, of the RNA and the magnitude of change is proportional to the number of base pair formations. Therefore, folding energy is used to characterize the local complementarity of nucleic acids.

I used a nucleic acid secondary structure predicting tool, mfold [10], to obtain d$G$ values and the lowest d$G$ was used in the calculations. Backtranslation and wobble base manipulations were performed by an online backtranslation tool [11]. The Backtranslation Tool (developed by Entelechon, Germany) creates DNA sequences from protein sequences using optional Genetic Code and Codon Usage Tables. Additionally, the user can determine special rules to select between available synonymous codons. I used this tool to create nucleic acid sequences where synonymous codons 1) were used accordingly to human CUT, 2) only the most frequent codons were used, 3) the codons were used with equal frequency (A=T=G=C in wobble positions) or 4) only A or C was used in wobble positions.

A JAVA tool called SeqForm, developed by us, was used to select sequence residues in predefined phases (every third in our case) and for residue replacements [12]. Amino acid collocations were detected and evaluated by another tool, SeqX [13].

Linear regression analyses and Student's *t*-tests were used for statistical analysis of the results.

## Results and discussion

*Folding free energy of coding sequences*

There is disagreement in the literature regarding the positional (free) folding energy of coding sequences. It has been shown that mRNAs have greater negative folding energies than shuffled or codon choice-randomized sequences [14], but this is not generally accepted [1]. I was able to find statistically significant dG differences between intact and shuffled mRNAs (Figure 1), but this FFE is only a small fraction (~10%) of the total folding energy. Also, there were many exceptions in this pool where the shuffled sequences had dG values equal to or even lower than those of the native sequences, which indicates that the primary structure (sequence) inhibits secondary structure formation in these cases.

The simple measurement of global folding energy is often criticized and is not regarded as a reliable measure of mRNA structure conservation [1]. There is compelling evidence for conserved, local secondary structures in the coding regions of eukaryotic mRNAs and pre-mRNAs [15] and widespread selection for local RNA secondary structure in coding regions of bacterial genes [16, 17].

Therefore I agree that global folding free energy measurements probably have little value in studies of single-stranded nucleic acid structures.

*The potential role of wobble basis in determining and regulating the folding energy of coding sequences*

Synonymous codons are not used with equal frequency in redundant codons; however, the frequency pattern of codon usage is well conserved in the same species. The cause or reason for this bias is not known, nor is it known whether the bias has any effect on biological functions. Many biological parameters are known to correlate with codon bias, for example (a) translational selection, (b) GC composition, (c) strand-specific mutational bias, (d) amino acid conservation, (e) protein hydropathy, (f) transcriptional selection, (g) structural elements in the coded protein, (h) tRNA copy number, (i) speed of translation and even (j) RNA stability or (k) gene length (for review see [18]) [19, 20]. It is logical to expect that codon usage bias should have some effect on mRNA structure and folding energies.

Indeed, modifications of synonymous codon usage have very pronounced effects on the dG values of coding sequences (Figure 1).

**Figure 1**

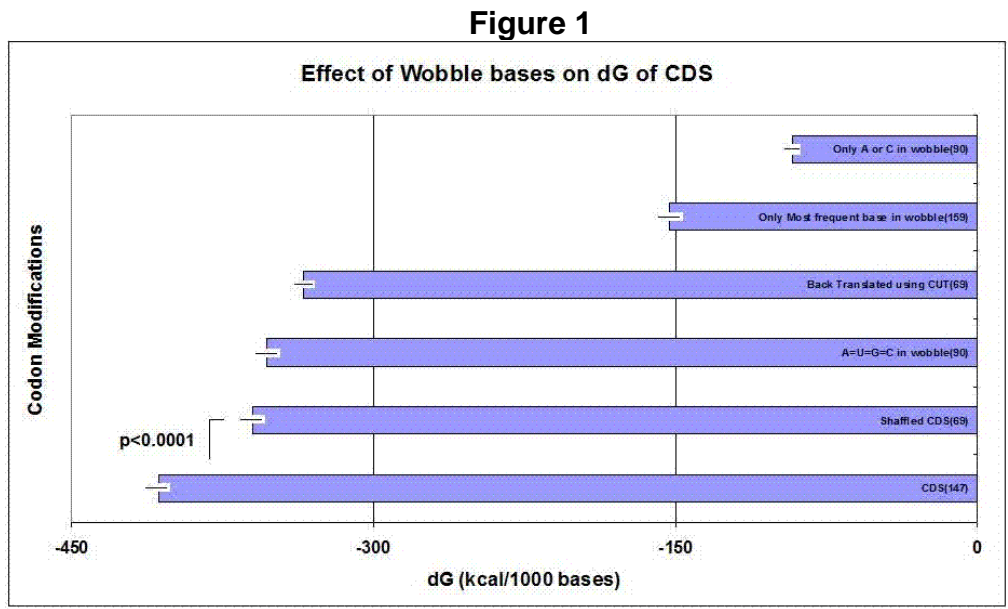

**Figure 1. Effect of wobble bases on the dG of CDS**
The TFE of mRNA is indicated in native sequences (CDS), after residue randomization (shuffle) and after the indicated manipulation of the wobble bases. (For details see the text). Each column represent the mean ± S.E.M.; n is indicated in the columns.

The TFE of our sequences (100%) was reduced to 88% by shuffling (all nucleotides, not only the third), to 87% by equalization of wobble base usage frequencies, and to 82% by back-translation of the protein sequence using the uniformly human Codon Usage Table (CUT). Much greater dG reduction was achieved when only the most frequent wobble bases were permitted (reduction to 38% of the intact mRNA) or when only the bases A or C were permitted in the wobble position (reduction to 22% of the intact molecule). It is possible to accomplish four-fold changes in mRNA folding energy by changing only the wobble bases, with no change in the primary sequence of the coded proteins. I conclude that the wobble bases are strong regulators of the total folding energy of mRNAs and probably even the mRNA structure.

The literature also suggests, along with our previous results [21], that there is a connection between the 3D structures of mRNA and the protein it encodes. This possibility makes the idea of wobble base regulation of nucleic acid structure even more interesting, because it indicates that the excess information in the redundant codon (carried by the wobble bases) may be used to modify or regulate the protein structure. To explore this possibility we compared the TFEs of intact and wobble base-modified mRNA sequences to the propensity towards different structural elements in the coded proteins (Figure 2).

**Figure 2**

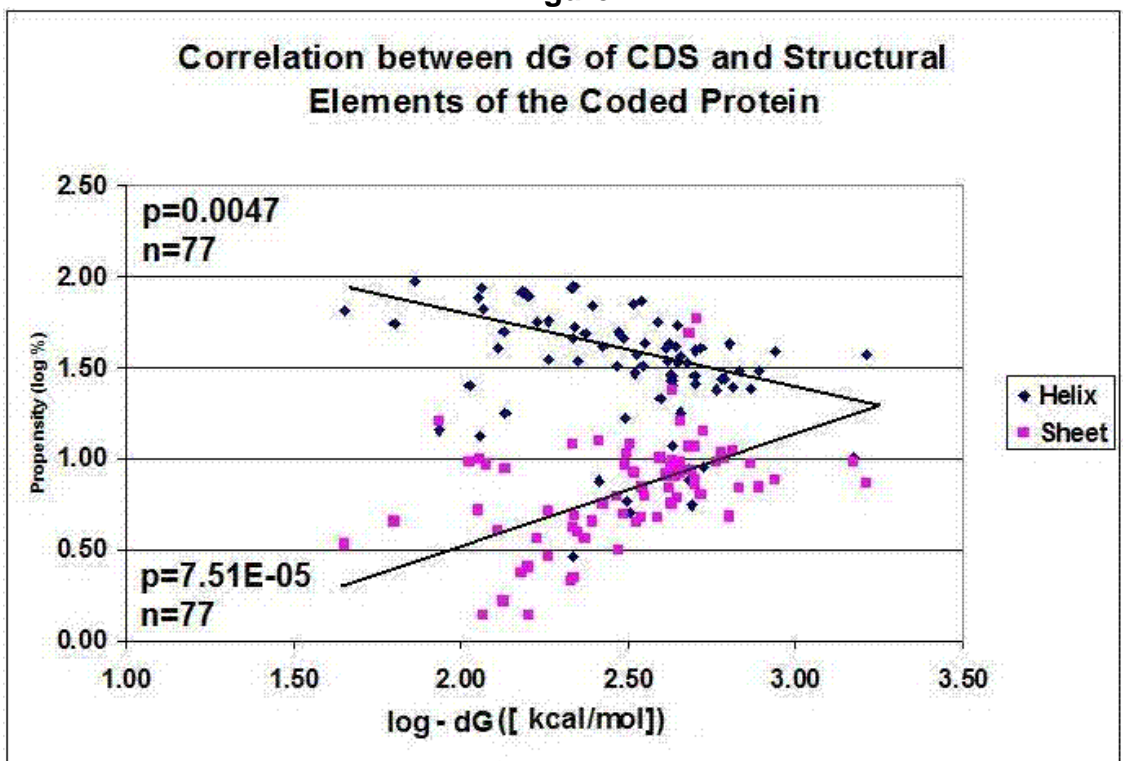

**Figure 2. Correlation between dG of CDS and Structural Elements of the Coded Protein.** The relative frequencies of the two main structural elements in 77 proteins are plotted against the folding energy of their coding sequences.

There is a positive correlation between mRNA TFE and the frequency of helices in the coded proteins. In contrast, the correlation between dG and beta sheet frequency is negative. This means that RNA complexity is proportional to beta sheet-type and other amino acid collocations, but inversely related to alpha helix frequency. The relationship between mRNA and protein structure will definitely be the subject of further evaluation because of its fundamental importance for further understanding the translation process and protein folding. The correlation between nucleic acid composition (sequence), nucleic acid folding energy (structure) and protein structural elements (helix, sheet) is a strong indication
that protein-folding information is present in the redundant exon bases (8, 21), and coding sequences function as chaperons (7).

*Correlation between mRNA sequence composition and total folding energy*

There is a strong, negative, linear correlation between the length (L) of a protein coding sequence (CDS) and its dG as well as its G+C (guanine + cytosine) content and folding energy (Figure 3).

**Figure 3**

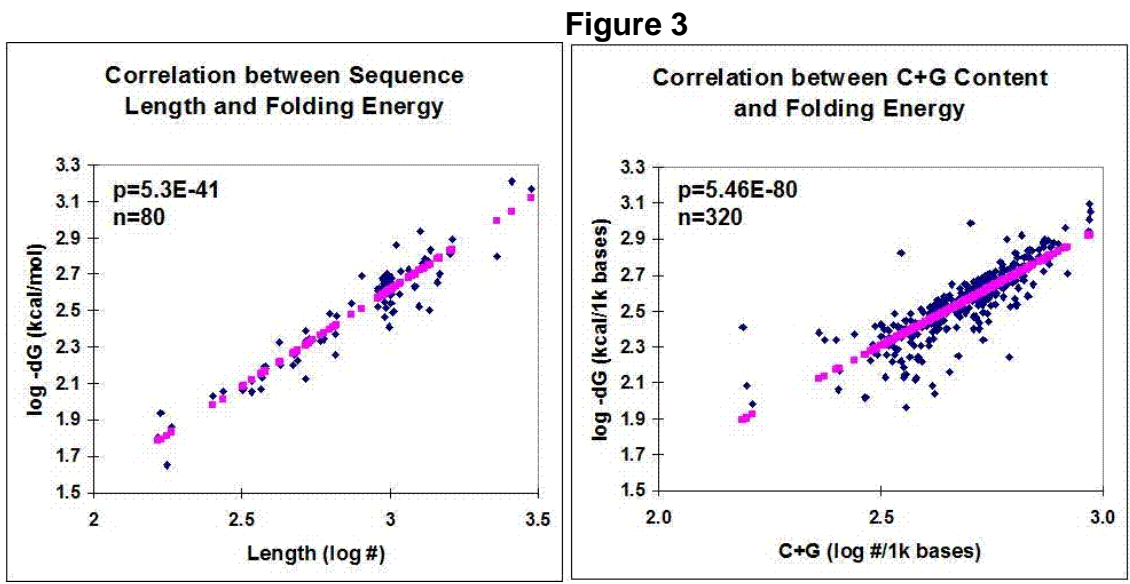

**Figure 3. Correlation between the Length (a), G+C content (b) and the TFE of mRNAs.**

C+G content is known to have major influence on the TFE of a nucleic acid [22]. The total C+G content varied from 27 to 70% in our sequence collection (50.3 ± 8.2, mean ± SD, n=80). The distribution of C+G differed between the different codon positions: more C+G was found in the first and third codon positions than in the second. Consequently, the FEs between RNA subsequences comprising the first and third codon positions were significantly lower than those involving the subsequences formed by 2$^{nd}$ codon residues (Figure 4).

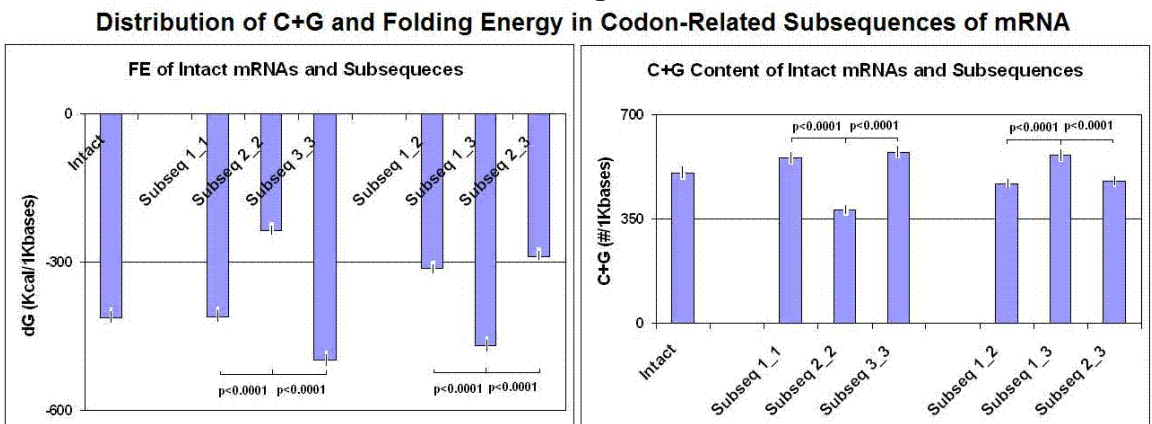

**Figure 4. Distribution of C+G and Folding Energy in Codon-Related Subsequences of mRNA**

Codons were identified in intact mRNA sequences and phase-selected for subsequences containing only the first (1_1), second (2_2), third (3_3) codon letters or combinations of these subsequences. Subsequence 1_2 means, for example, that the 3$^{rd}$ codon letters were removed from the original mRNA. The G+C content was counted and the dGs were measured by *mfold*. Each column represent the mean ± S.E.M of n=87 determinations.

The TFE represents the 3D complexity of the structures formed by nucleic acids where Watson-Crick base pairs play a fundamental role. There are 3 H bonds between C and G (dG = -1524 kcal/1000 bases) but only two between A and T (dG = -365 kcal/1000 bases). This more than 4-fold dG difference explains the dominance of C+G over A+T in determining TFE.

Two major factors determine the folding energies of nucleic acids: first, the G+C content or (G+C)/(A+T) ratio; second, the availability of bases for Watson-Crick base pairing, which may be characterized by the 1000-2|G-C|-2|A-T| values. The |G-C| and |A-T| values are the absolute difference between WC pairs, i.e. the non-pairing fraction. The value 1000 – (non-pairing fractions) will give the maximum possible number of WC pairs in a 1000 base-long sequence.

There are strong linear correlations between these values (based on the primary sequence of the mRNA) and the free energy (Figure 5). The correlation is strong enough to be able to predict the dG value of a nucleic acid from the base composition of its primary sequence. This indicates that the relationship between nucleic acid sequence composition and folding energy is simple. I used coding sequences in this experiment, but this relationship is not unique to CDS; it is generally valid for all kinds of nucleic acids. The nucleotides at the wobble positions have a large influence on the folding energy of mRNAs. Synonymous codons are seemingly equivalent to each other in respect of amino acid coding, but they are not at all synonymous or freely interchangeable with regard to mRNA folding energy and secondary structure.

**Figure 5**

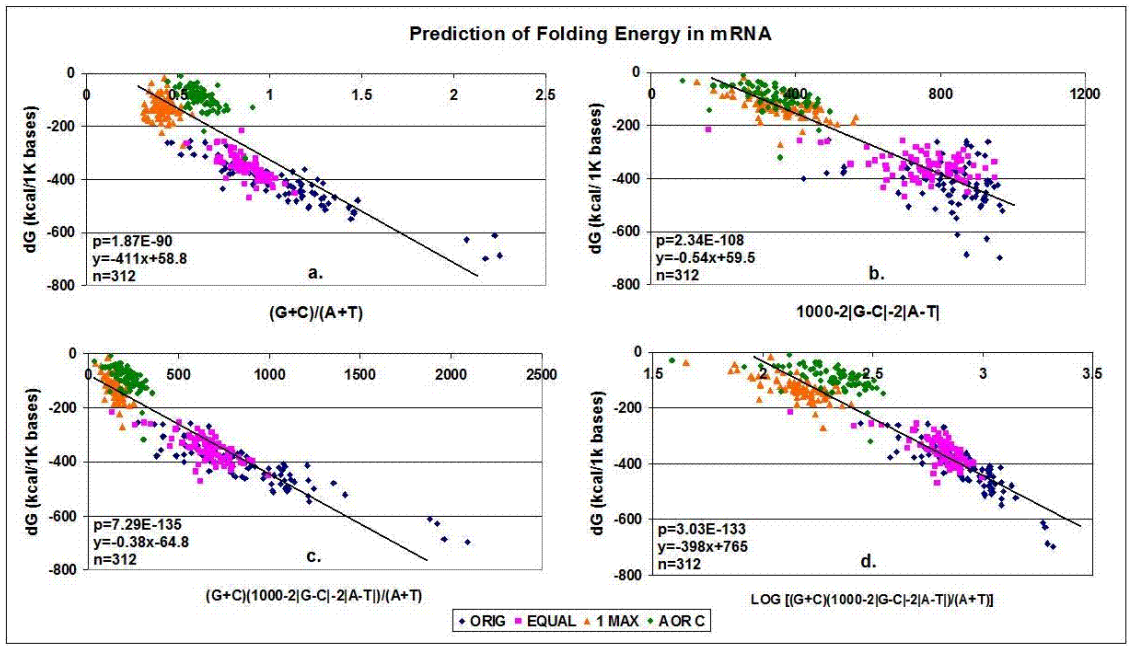

**Figure 5. Correlation between nucleotide composition and Folding Energy in mRNA sequences.**

The wobble base compositions of 78 mRNA sequences (ORIG) were modified to contain equal numbers of synonymous codons (EQUAL); only one, the most frequent synonymous codon (1 MAX); or A or C, but not T or G (A OR C). Predictions of dG were made using the different equations in the figure and plotted against the *mfold* values determined.

*Contribution of synonymous codons to the folding energy of coding sequences*

To analyze the role of individual codons on the mRNA structure further, I designed 64 different artificial nucleic acid sequences. Each sequence contained a 100-fold repetition of a single codon, i.e. they were repeating poly-codons. The sequences (64x64) were hybridized using the DINAMelt Server two-stage hybridization tool [23]. The 4096 codon affinity (dG) values were sorted into 400 subgroups corresponding to the 20x20 coded amino acid pairs. The 400 accumulated (summed) codon affinity values are given in Table I.

Fifty-one of the 400 possible amino acid pairs are coded by codons with no affinity for each other. The accumulated dG values of the codons encoding the remaining 349 amino acid pairs show statistically highly significant positive correlations to the calculated and real frequencies of amino acid pairs in real proteins (Figure 6).

# Table I

## Accumulated Affinity (dG) between Amino Acid Codons

| CF | 1 | 1 | 2 | 2 | 2 | 2 | 2 | 2 | 2 | 2 | 2 | 3 | 3 | 4 | 4 | 4 | 4 | 4 | 6 | 6 | 6 |
|---|---|---|---|---|---|---|---|---|---|---|---|---|---|---|---|---|---|---|---|---|---|
| AA | Met | Trp | Lys | Glu | Phe | Asn | Gln | His | Tyr | Asp | Cys | End | Ile | Pro | Gly | Thr | Ala | Val | Leu | Arg | Ser |
| 1  Met | 2.8 | 2.6 | 0.0 | 0.0 | 3.2 | 3.2 | 3.3 | 3.6 | 3.1 | 2.9 | 3.1 | 3.2 | 3.7 | 3.4 | 2.8 | 3.4 | 3.2 | 3.6 | 3.7 | 3.4 | 3.9 |
| 1  Trp | 2.6 | 1.0 | 0.0 | 0.0 | 3.1 | 3.2 | 3.3 | 3.7 | 3.6 | 3.2 | 3.4 | 2.9 | 3.5 | 3.9 | 3.0 | 3.9 | 3.8 | 3.0 | 4.0 | 3.7 | 3.7 |
| 2  Lys | 0.0 | 0.0 | 0.0 | 0.0 | 3.8 | 0.0 | 0.0 | 3.1 | 3.3 | 0.0 | 3.3 | 0.0 | 3.4 | 3.1 | 0.0 | 2.9 | 2.9 | 3.4 | 3.9 | 3.1 | 3.9 |
| 2  Glu | 0.0 | 0.0 | 0.0 | 0.0 | 3.9 | 0.0 | 0.0 | 3.6 | 3.4 | 0.0 | 3.4 | 0.0 | 3.5 | 3.9 | 0.0 | 3.5 | 3.5 | 3.5 | 4.2 | 3.6 | 4.2 |
| 2  Phe | 3.2 | 3.1 | 3.8 | 3.9 | 0.0 | 3.4 | 3.3 | 0.0 | 0.0 | 3.5 | 0.0 | 3.6 | 3.1 | 0.0 | 4.0 | 3.5 | 3.3 | 3.3 | 0.0 | 4.0 | 3.2 |
| 2  Asn | 3.2 | 3.2 | 0.0 | 0.0 | 3.4 | 2.9 | 0.0 | 2.8 | 3.5 | 3.2 | 3.7 | 3.6 | 3.6 | 0.0 | 3.2 | 2.9 | 3.0 | 3.9 | 4.0 | 3.2 | 3.8 |
| 2  Gln | 3.3 | 3.3 | 0.0 | 0.0 | 3.3 | 0.0 | 3.2 | 0.0 | 3.2 | 3.3 | 4.0 | 3.6 | 2.8 | 3.3 | 3.5 | 2.9 | 3.9 | 4.0 | 4.1 | 3.7 | 3.9 |
| 2  His | 3.6 | 3.7 | 3.1 | 3.6 | 0.0 | 2.8 | 0.0 | 2.7 | 2.7 | 3.7 | 3.6 | 3.8 | 3.2 | 0.0 | 4.0 | 3.1 | 3.6 | 4.0 | 3.7 | 3.9 | 3.7 |
| 2  Tyr | 3.1 | 3.6 | 3.3 | 3.4 | 0.0 | 3.5 | 3.2 | 2.7 | 3.5 | 3.2 | 3.2 | 3.9 | 3.7 | 0.0 | 3.9 | 3.4 | 3.3 | 4.0 | 3.7 | 3.8 | 4.0 |
| 2  Asp | 2.9 | 3.2 | 0.0 | 0.0 | 3.5 | 3.2 | 3.3 | 3.7 | 3.2 | 3.2 | 3.5 | 3.4 | 3.8 | 3.7 | 3.4 | 3.5 | 3.5 | 4.0 | 4.0 | 3.9 | 4.2 |
| 2  Cys | 3.1 | 3.4 | 3.3 | 3.4 | 0.0 | 3.7 | 4.0 | 3.6 | 3.2 | 3.5 | 3.5 | 3.7 | 3.6 | 3.6 | 4.0 | 3.9 | 4.2 | 3.8 | 3.7 | 4.1 | 3.9 |
| 3  End | 3.2 | 2.9 | 0.0 | 0.0 | 3.6 | 3.6 | 3.6 | 3.8 | 3.9 | 3.4 | 3.7 | 3.8 | 4.0 | 3.7 | 3.3 | 3.9 | 3.8 | 3.9 | 4.3 | 3.8 | 4.2 |
| 3  Ile | 3.7 | 3.5 | 3.4 | 3.5 | 3.1 | 3.6 | 2.8 | 3.2 | 3.7 | 3.8 | 3.6 | 4.0 | 3.9 | 0.0 | 3.8 | 3.6 | 3.2 | 4.0 | 4.0 | 3.8 | 4.0 |
| 4  Pro | 3.4 | 3.9 | 3.1 | 3.9 | 0.0 | 0.0 | 3.3 | 0.0 | 0.0 | 3.7 | 3.6 | 3.7 | 0.0 | 3.2 | 4.5 | 3.3 | 4.1 | 4.2 | 3.6 | 4.3 | 3.9 |
| 4  Gly | 2.8 | 3.0 | 0.0 | 0.0 | 4.0 | 3.2 | 3.5 | 4.0 | 3.9 | 3.4 | 4.0 | 3.3 | 3.8 | 4.5 | 3.5 | 4.2 | 4.3 | 4.0 | 4.5 | 4.2 | 4.4 |
| 4  Thr | 3.4 | 3.9 | 2.9 | 3.5 | 3.5 | 2.9 | 2.9 | 3.1 | 3.4 | 3.5 | 3.9 | 3.9 | 3.6 | 3.3 | 4.2 | 3.2 | 3.9 | 4.4 | 4.2 | 4.2 | 4.3 |
| 4  Ala | 3.2 | 3.8 | 2.9 | 3.5 | 3.3 | 3.0 | 3.9 | 3.6 | 3.3 | 3.5 | 4.2 | 3.8 | 3.2 | 4.1 | 4.3 | 3.9 | 4.5 | 4.4 | 4.3 | 4.4 | 4.3 |
| 4  Val | 3.6 | 3.0 | 3.4 | 3.5 | 3.3 | 3.9 | 4.0 | 4.0 | 4.0 | 4.0 | 3.8 | 3.9 | 4.0 | 4.2 | 4.0 | 4.4 | 4.4 | 3.6 | 4.3 | 4.4 | 4.1 |
| 6  Leu | 3.7 | 4.0 | 3.9 | 4.2 | 0.0 | 4.0 | 4.1 | 3.7 | 3.7 | 4.0 | 3.7 | 4.3 | 4.0 | 3.6 | 4.5 | 4.2 | 4.3 | 4.3 | 4.0 | 4.5 | 4.3 |
| 6  Arg | 3.4 | 3.7 | 3.1 | 3.6 | 4.0 | 3.2 | 3.7 | 3.9 | 3.8 | 3.9 | 4.1 | 3.8 | 3.8 | 4.3 | 4.2 | 4.2 | 4.4 | 4.4 | 4.5 | 4.5 | 4.6 |
| 6  Ser | 4.0 | 3.8 | 3.9 | 4.2 | 3.2 | 3.7 | 3.9 | 3.5 | 3.8 | 4.2 | 4.0 | 4.2 | 3.8 | 3.8 | 4.5 | 4.2 | 4.4 | 4.2 | 4.3 | 4.6 | 4.1 |

Each value is the accumulated affinity (dG) between synonymous codons corresponding to the indicated amino acids (log -kcal/mol);
CF: codon frequency / number of synonymous codons/amino acid. Highest (red) and lowest (blue) values are color coded.

# Figure 6

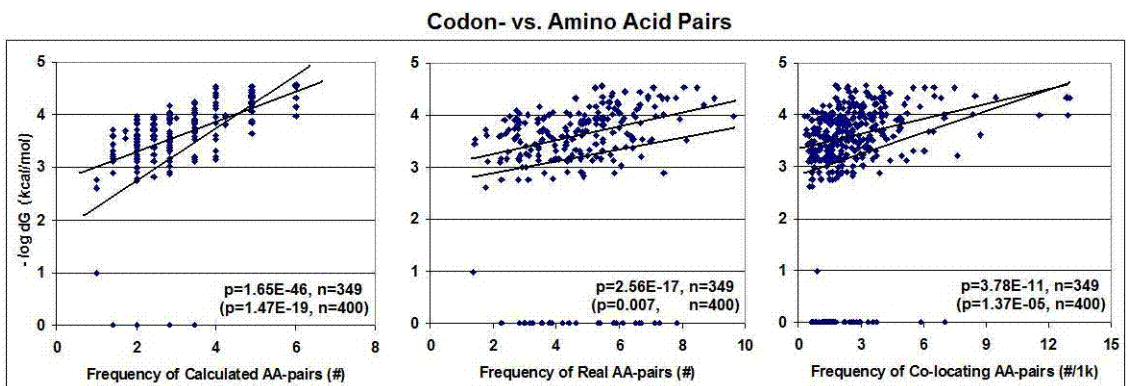

**Figure 6. Correlation between the affinity (dG) of artificial poly-codons and the propensity for the coded amino acid pairs.**

The affinities of artificial poly-codons for each other (64x64=4096 dG) were determined. Synonymous codons were combined and added to 20x20=400 subgroups corresponding to the possible pairs formed by the respective coded amino acids. Frequency of calculated amino acid (AA)-pairs= sqrt [(# of synonymous codon of amino acid A)x(# of synonymous codon of amino acid B)]. Frequencies of real AA-pairs were calculated from the real propensity towards amino acids in the proteins examined using the equation = sqrt [(propensity of amino acid A)x(propensity of amino acid B)]. Frequencies of co-locating AA-pairs were measured from crystallographic structures using the SeqX tool. Two linear correlation coefficients were calculated, one excluding and the other () including dG=0 values.

This chain of correlations indicates that synonymous codon frequency has a strong positive effect on the accumulated affinity of codons in the nucleic acid (mRNA structure) and the relative frequencies of amino acids in the coded protein generally and co-locating amino acids especially (protein structure). I interpret this result as supporting the view that the effects of synonymous codons on nucleic acid structure and protein composition are additive and they are not interchangeable with each other in these respects.

An additional chain of evidence that suggests the non-equality of synonymous codons arises from studies on single-nucleotide polymorphisms (SNPs). Synonymous SNPs do not produce altered coding sequences, so they are not expected to change the function of the protein in which they occur. However, they often do. Synonymous SNPs in the Multi-drug Resistance 1 (MDR1) gene alter the conformation of the protein product, the P-glycoprotein (P-gp), while the mRNA and protein levels remain similar in wild-type and polymorphic P-gp [24, 25]. Synonymous mutations in the human dopamine receptor D2 (DRD2) affect mRNA stability and synthesis of the receptor [26]. Synonymous mutations affect splicing and are not neutral in evolution [27,28]. After all, SNPs are not silent and not invisible in many cases [29, 30]. It might be hypothesized that the presence of a rare codon, marked by the synonymous polymorphism, affects the timing of co-translational folding and thereby alters the structure and function of expressed protein. My recent study provides additional evidence in that direction.

There are numerous theories and speculations regarding the Genetic Code because of its redundancy. The 64>20 canonical code contains 3-times more information than it is necessary for determining amino acids. The excess information is stored in the wobble bases. A useful property of the Genetic Code is the minimization of the effects of frame-shift translation errors. However it is more and more accepted that coding sequences convey, in addition to the protein-coding information, several different biologically meaningful signals at the same time. These "parallel codes" include binding sequences for regulatory 31-34] and structural proteins [35], signals for splicing [36], and RNA secondary structure [16, 37-39] It was recently noted [40] that the universal genetic code can allow arbitrary sequences of nucleotides within coding regions much better than the vast majority of other possible genetic codes. This ability to support parallel codes is strongly correlated with an additional property—minimization of the effects of frame-shift translation errors.

My study at hand is a specific case where the folding free energy of an mRNA is used as a measure for adapting structural features and the suggestions that wobble bases are important in carrying parallel (structural) codes are not entirely novel. However it is an additional indication that the interchangeability of synonymous codons is rather restricted: the universal genetic code is optimal as it is.

The reported correlations are not backed up with an evolutionary model and doesn't show that the RNA secondary structure has exerted a selection pressure on the evolution of the canonical genetic code. Therefore these correlations are not necessarily genuine and they may simply be secondary by-product of other optimization processes. However, the existence of these correlations *per se* provide strong support for the presence and importance of parallel codes in the universal, canonical codon.

## Conclusion

The genetic code is redundant in regard to the meaning of codons in translation. However, this redundancy seems to have its own meaning. Synonymous codons are interchangeable in regard to amino acid coding, but their effect is individual and additive in respect of their role in determining the FEs of coding sequences and the relative frequencies of amino acids in the translated protein sequences and co-locating amino acid pairs. This might explain why wobble base modifications often have a large effect on the efficiency of gene expression and folding of the protein product [41-48].

## Competing interests

The author declares that he has no competing interests.

## Acknowledgements

The continuous support and editorial help of Dr P S Agutter is very much acknowledged.

## References

1. Workman C, Krogh A: **No evidence that mRNAs have lower folding free energies than random sequences with the same dinucleotide distribution**. *Nucleic Acids Res* 2002, **24**:4816-4822.

2. Patterson DJ, Yasuhara K, Ruzzo WL: **PRE-mRNA Secondary Structure Prediction Aids, Splice Site Prediction.** *Pacific Symposium on Biocomputing* 2002, **7**:223-234.

3. Winkler WC, Cohen-Chalamish S, Breaker RR: **An mRNA structure that controls gene expression by binding FMN.** *Proc Natl Acad Sci U S A* 2002, **99**:15908-15913.

4. Mita K, Ichimura S, Zama M, James TC: **Specific codon usage pattern and its implications on the secondary structure of silk fibroin mRNA.** *J Mol Biol 1988*, **203**:917–925.

5. Zama M: **Correlation between mRNA structure of the coding region and translational pauses**. *Nucleic Acids Symp Ser* 1999, **42**:81–82.

6. Jia M, Luo L, Liu C: **Statistical correlation between protein secondary structure and messenger RNA stem–loop structure.** *Biopolymers* 2004, **73**:16–26.

7. Biro JC: **Nucleic acid chaperons: a theory of an RNA-assisted protein folding.** *Theor Biol Med Model* 2005, **2**:35.

8. Biro JC: **Indications that "codon boundaries" are physico-chemically defined and that protein-folding information is contained in the redundant exon bases.** *Theor Biol Med Model* 2006, **3**:28.

9. Shaper EG: **The secondary structure of mRNAs from Escherichia coli: its possible role in increasing the accuracy of translation.** *Nucleic Acids Res* 1985, **13**:275-288.

10. Zuker M: **Mfold web server for nucleic acid folding and hybridization prediction.** *Nucleic Acids Res* 2003, **31**:3406-3415.

11. **Backtranslation Tool**, *Entelechon*, http://www.entelechon.com/bioinformatics/backtranslation.php?lang=eng

12. Biro JC: **SeqForm.** www.janbiro.com/downloads 2005.

13. Biro JC, Fordos G: **SeqX: a tool to detect, analyze and visualize residue co-locations in protein and nucleic acid structures.** *BMC Bioinformatics* 2005, **12**:170.

14. Seffens W, Digby D: **mRNAs have greater negative folding free energies than shuffled or codon choice randomized sequences.** *Nucleic Acids Res* 1999, **27**:1578–1584.

15. Meyer IM, Miklós I: **Statistical evidence for conserved, local secondary structure in the coding regions of eukaryotic mRNAs and pre-mRNAs.** *Nucleic Acids Res* 2005, **33**:6338-6348.

16. Katz L, Burge CB: **Widespread selection for local RNA secondary structure in coding regions of bacterial genes.** *Genome Res* 2003, **13**:2042-2051.

17. Shabalina SA, Ogurtsov AY, Spiridonov NA: **A periodic pattern of mRNA secondary structure created by the genetic code.** *Nucleic Acids Research* 2006, **34**:2428–2437.

18. Ermolaeva MD: **Synonymous codon usage in bacteria.** *Curr Issues Mol Biol* 2001, **3**:91-97.

19. Powell JR, Moriyama EN: **Evolution of codon usage bias in Drosophila.** *Proc Natl Acad Sci U S A.* 1997, **94:**7784-7790.

20. Chamary JV, Hurst LD: **Evidence for selection on synonymous mutations affecting stability of mRNA secondary structure in mammals.** *Genome Biology* 2005, **6**:R75.

21. Biro JC: **Protein folding information in nucleic acids which is not present in the genetic code.** *Ann N Y Acad Sci* 2006, **1091**:399-411.

22. Seffens W: **mRNA Classification Based on Calculated Folding Free Energies**. *The World Wide Web Journal of Biology* 1999, http://epress.com/w3jbio/vol4/seffens/index.html

23. Markham NR, Zuker M: **DINAMelt Servers two-stage hybridization tool**, 2005. http://www.bioinfo.rpi.edu/applications/hybrid/twostate.php)

24. Kimchi-Sarfaty C, Oh JM, Kim IW, Sauna ZE, Calcagno AM, Ambudkar SV, Gottesman MM: **A "silent" polymorphism in the MDR1 gene changes substrate specificity.** *Science* 2007, **315**:525-528.

25. Sauna ZE, Kimchi-Sarfaty C, Ambudkar SV, Gottesman MM: **The sounds of silence: synonymous mutations affect function.** *Pharmacogenomics* 2007, **8**:527-532.

26. Duan J, Wainwright MS, Comeron JM, Saitou N, Sanders AR, Gelernter J, Gejman PV: **Synonymous mutations in the human dopamine receptor D2 (DRD2) affect mRNA stability and synthesis of the receptor.** *Hum Mol Genet* 2003, **12**:205-216.

27. Pagani F, Raponi M, Baralle FE: **Synonymous mutations in CFTR exon 12 affect splicing and are not neutral in evolution.** *Proc Natl Acad Sci U S A* 2005, **102**:6368-6372.

28. Nielsen KB, Sorensen S, Cartegni L, Corydon TJ, Doktor TK, Schroeder LD, Reinert LS, Elpeleg O, Krainer AR, Gregersen N, Kjems J, Andresen BS: **Seemingly neutral polymorphic variants may confer immunity to splicing-inactivating mutations: a synonymous SNP in exon 5 of MCAD protects from deleterious mutations in a flanking exonic splicing enhancer.** *Am J Hum Genet* 2007, **80**:416-432. Erratum in: *Am J Hum Genet* 2007, **80**:816.

29. Komar AA: **Genetics. SNPs, silent but not invisible.** *Science* 2007, **315**:466-467.

30. Soares C: **Codon spell check. Silent mutations are not so silent after.** *Sci Am* 2007, 296:23-24.

31. Robison K, McGuire AM, Church GM: **A comprehensive library of DNA-binding site matrices for 55 proteins applied to the complete *Escherichia coli* K-12 genome.** *J. Mol. Biol.* 1998, **284:**241–254.

32. Stormo GD: **DNA binding sites: Representation and discovery.** *Bioinformatics* 2000, **16:**16–23.

33. Lieb JD, Liu X, Botstein D, Brown PO: **Promoter-specific binding of Rap1 revealed by genome-wide maps of protein-DNA association.** *Nat. Genet.* 2001, **28:** 327–334.

34. Kellis M, Patterson N, Endrizzi M, Birren B, Lander ES: **Sequencing and comparison of yeast species to identify genes and regulatory elements.** *Nature* 2003, **423:**241–254.

35. Draper DE: **Themes in RNA-protein recognition.** *J. Mol. Biol.* 1999, **293:**255–270.

36. Cartegni L, Chew SL, Krainer AR: **Listening to silence and understanding nonsense: Exonic mutations that affect splicing.** *Nat. Rev. Genet.* 2002, **3:**285–298.

37. Zuker M, Stiegler P: **Optimal computer folding of large RNA sequences using thermodynamics and auxiliary information.** *Nucleic Acids Res.* 1981, **9:**133–148

38. Shpaer EG: **The secondary structure of mRNAs from *Escherichia coli*: Its possible role in increasing the accuracy of translation.** *Nucleic Acids Res.* 1985, **13:**275–288.

39. Konecny J, Schoniger M, Hofacker I, Weitze MD, Hofacker GL: **Concurrent neutral evolution of mRNA secondary structures and encoded proteins.** *J. Mol. Evol.* 2000, **50:**238–242.

40. Itzkovitz S, Alon U : **The genetic code is nearly optimal for allowing additional information within protein-coding sequences.** *Genome Res.* 2007, **17**:405-412.

41. Mukhopadhyay P, Basak S, Ghosh TC: **Synonymous codon usage in different protein secondary structural classes of human genes: implication for increased non-randomness of GC3 rich genes towards protein stability.** *J Biosci.* 2007, **32**:947-63.

42. Gu W, Zhou T, Ma J, Sun X, Lu Z: **The relationship between synonymous codon usage and protein structure in Escherichia coli and Homo sapiens.** *Biosystems.* 2004, **73**:89-97.

43. Gu W, Zhou T, Ma J, Sun X, Lu Z: **Folding type specific secondary structure propensities of synonymous codons.** *IEEE Trans Nanobioscience.* 2003, **2**:150-157.

44. Oresic M, Dehn M, Korenblum D, Shalloway D: **Tracing specific synonymous codon-secondary structure correlations through evolution.** *J Mol Evol.* 2003, **56**:473-84.

45. Cortazzo P, Cerveñansky C, Marín M, Reiss C, Ehrlich R, Deana A: **Silent mutations affect in vivo protein folding in Escherichia coli.** *Biochem Biophys Res Commun.* 2002, **293**:537-541.

46. Komar AA, Lesnik T, Reiss C: **Synonymous codon substitutions affect ribosome traffic and protein folding during in vitro translation.** *FEBS Lett.* 1999, **462**:387-91.

47. Xie T, Ding D: **The relationship between synonymous codon usage and protein structure.** FEBS Lett. 1998 Aug 28;434(1-2):93-6. **Erratum in:** FEBS Lett 1998, **437**:164. Tao, X [corrected to Xie, T]; Dafu, D [corrected to Ding, D].

48. Oresic M, Shalloway D: **Specific correlations between relative synonymous codon usage and protein secondary structure.** *J Mol Biol.* 1998, **281**:31-48.